\documentclass[final]{raa06}           
\usepackage{graphicx,times}             
\usepackage{natbib}
\usepackage{amssymb,amsmath}
\bibpunct{(}{)}{;}{a}{}{,}
\usepackage[colorlinks=true, citecolor=blue]{hyperref}%

\usepackage{graphicx,kantlipsum,setspace}
\usepackage{caption}
\usepackage{graphics,epsf}
\usepackage{amsmath}                
\usepackage{amsfonts}               
\usepackage{amssymb}                
\usepackage{epsfig}                 
\usepackage{appendix}
\usepackage{graphicx}
\usepackage{float}
\usepackage{color}
\usepackage{multirow}
\usepackage{colortbl}
\usepackage[para,online,flushleft]{threeparttable}
\usepackage{xcolor}

\hypersetup{citecolor=blue, 
            linkcolor=red, 
            menucolor=blue, 
            urlcolor=blue}  

\newcommand{\cm}{{~\rm cm}}

\newcommand{\km}{{~\rm km}}
\newcommand{\s}{{~\rm s}}

\newcommand{\g}{{~\rm g}}

\newcommand{\yr}{{~\rm yr}}

\newcommand{\AU}{{~\rm AU}}

\begin{document}

   \title{A pre-explosion effervescent zone for the circumstellar material in SN~2023ixf
}

   \volnopage{Vol.0 (20xx) No.0, 000--000}      
   \setcounter{page}{1}          

   \author{Noam Soker
    }

   \institute{Department of Physics, Technion, Haifa, 3200003, Israel;   {\it   soker@physics.technion.ac.il}\\
\vs\no
   {\small Received~~20xx month day; accepted~~20xx~~month day}}

\abstract{I present the effervescent zone model to account for the compact dense circumstellar material (CSM) around  the progenitor of the core collapse supernova (CCSN) SN~2023ixf. The effervescent zone is composed of bound dense clumps that are lifted by stellar pulsation and envelope convection to distances of $\approx {\rm tens} \times \AU$, and then fall back. The dense clumps provide most of the compact CSM mass and exist alongside the regular (escaping) wind. I crudely estimate that for a compact CSM within $R_{\rm CSM} \approx 30 \AU$ that contains $M_{\rm CSM}  \approx 0.01 M_\odot$, the density of each clump is $k_b \ga 3000$ times the density of the regular wind at the same radius and that the total volume filling factor of the clumps is several percent. The clumps might cover only a small fraction of the CCSN photosphere in the first days post-explosion, accounting for the lack of strong narrow absorption lines. The long-lived effervescent zone is compatible with no evidence for outbursts in the years prior to SN~2023ixf explosion and the large-amplitude pulsations of its progenitor, and it is an alternative to the CSM scenario of several-years-long high mass loss rate wind.  
}

\authorrunning{N. Soker}            
\titlerunning{An effervescent zone in SN~2023ixf}  
   
      \maketitle

\section{Introduction}
\label{sec:intro}

A number of studies conclude that the ejecta of the type II core collapse supernova (CCSN) SN~2023ixf interacted with a circumstellar material (CSM) that was extended up to a distance of $R_{\rm CSM} \simeq 20-50 \AU$ (e.g., \citealt{Bergeretal2023,  Bostroemetal2023, Grefenstetteetal2023, JacobsonGalanetal2023, Kilpatricketal2023, SinghTejaetal2023, SmithNetal2023}).  

Two basic types of models might account for a compact CSM. In one the red supergiant (RSG) progenitor of a CCSN with a close CSM experiences a pre-explosion high-mass ejection episode that starts years to weeks before explosion (e.g., \citealt{Foleyetal2007, Pastorelloetal2007, Smithetal2010, Marguttietal2014, Ofeketal2014, SvirskiNakar2014, Tartagliaetal2016, Yaronetal2017, Wangetal2019, Bruchetal2020, Prenticeetal2020, Strotjohannetal2021}). Most of the studies of the CSM of SN~2023ixf take it to result from an outburst that took place few years before explosion, e.g., \cite{Bostroemetal2023} take the wind velocity to be $v_{\rm w} \simeq 10 \km \s^{-1}$ and deduce the beginning of the high mass loss rate to be $\tau_{\rm w} \simeq 17 \yr$ before explosion while \cite{JacobsonGalanetal2023} take $v_{\rm w} \simeq 50 \km \s^{-1}$ to estimate $\tau_{\rm w} \simeq 3-6 \yr$. \cite{SmithNetal2023} observationally infer the expansion velocity of the CSM to be $v_{\rm w}=115 \km \s^{-1}$ and from that they calculate $\tau_{\rm w} \simeq 0.9-1.5 \yr$. \cite{SmithNetal2023} consider the CSM to be non-spherical.   

The high mass loss rate phase shortly before explosion might be powered by a binary interaction that is triggered by the sudden expansion of the RSG star years to weeks before explosion (e.g., \citealt{Soker2013PEO, SmithArnett2014})\footnote{\cite{Soker2013PEO} suggested (based on earlier binary models of similar types of transients) that a sudden pre-CCSN swelling may trigger asymmetric mass loss in binary systems. Five months later \cite{SmithArnett2014} repeated this suggestion. Later papers, e.g., \cite{Smith2017} and \cite{SmithNetal2023}, wrongly attributed this suggestion to \cite{SmithArnett2014}.}. The binary interaction likely involves jets (e.g., a review by \citealt{Soker2022jets}). A  problem with this type of CSM is that there was no recorded outburst of the progenitor of SN~2023ixf. \cite{Neustadtetal2023} find no outbursts in the time period of 5600 to 463 days before explosion. \cite{Jencsonetal2023} find no evidence for an outburst, beside the periodic variability, from 2010 until 10 days before explosion. \cite{Soraisametal2023} find that the star continues its general variability over 16 years until less than two weeks before explosion. 

The second possibility for a compact CSM, up to $\approx 100 \AU$, is a long-lived extended material around the CCSN progenitor. \cite{Dessartetal2017} presented a model of a long-lived complex extended dense zone of $\approx 0.01 M_\odot$ around RSG progenitors of CCSNe. \cite{Moriyaetal2017} consider a dense compact CSM that is the acceleration zone of the wind (also \citealt{Moriyaetal2018}). 
In this study I consider a different extended material above the RSG progenitor of SN~2023ixf, namely, the \textit{effervescent CSM zone} model that I proposed to occur in some CCSN progenitors \citep{Soker2021effer}. 

The effervescent zone is an extended zone from the giant surface and up to $R_{\rm eff} \approx (10-100) R_\ast$ where in addition to the regular (escaping) stellar wind there are many dense clumps that do not reach the escape velocity, and hence rise and fall back.  The average density of the bound mass might reach tens of times that of the regular wind. Since the effervescent zone can live for thousands of years and more it removes the need for many type II CCSN progenitors to experience a very strong outburst just years to weeks before explosion.   
Namely, the long-lived effervescent zone can mimic a short-duration enhanced mass loss rate phase. Again, in many cases the RSG does experience an outburst before explosion, but not in all cases. 


The effervescent CSM model for RSG stars is based on  the effervescent zone model for asymptotic giant branch (AGB) stars \citep{Soker2008eff}. This model was motivated by observations of inhomogeneous outflows from AGB stars, including clumps (dusty or not) and sometimes with inflows alongside the outflow (e.g., \citealt{Lopezetal1997, Vlemmingsetal2002, DiamondKemball2003, Cottonetal2006, Fonfriaetal2008, KlochkovaChentsov2008, TuyetNhung2019, Khourietal2020}). A good example is the  pulsating AGB star Mira~A that has a radius of $R_\ast \simeq 500 R_\odot$ (e.g., \citealt{WoodKarovska2006}) and that possesses an inhomogeneous and clumpy asymmetrical compact CSM, up to $\approx {\rm several} \times 10 \AU$ (e.g., \citealt{Planesasetal1990, RydeSchoier2001, Lopezetal1997}). 

The presence of compact CSM around some CCSNe and observations of inhomogeneous winds, sometimes with clumps, and/or inflow/outflow in some RSGs (e.g., \citealt{LobelDupree2000, Humphreysetal2007, JosselinPlez2007, Ohnakaetal2011, Kervellaetal2016, Kaminski2019}) and claims for extended regions where inflow and outflow coexist around stars that are close to their Eddington luminosity limit (e.g., \citealt{OwockivanMarle2008, vanMarleetal2009}) motivated me to develop the effervescent zone model for CCSN progenitors (see \citealt{Soker2021effer} for more details). 

In this Letter I propose that the compact CSM around the progenitor of SN~2023ixf was an effervescent zone rather than the ejecta of a short-lived enhanced mass loss rate episode. I mostly motivated by the lack of indication of a pre-explosion outburst and from the large-amplitude pulsations of the progenitor. My main goal is to estimate plausible parameters for the effervescent zone (section \ref{sec:Effervescent}). I discuss the implications in section \ref{sec:Summary}.

\section{The pre-explosion effervescent zone}
\label{sec:Effervescent}

\subsection{An effervescent zone}
\label{subsec:EffervescentZone}

\begin{figure}  [t]
\centering
\begin{tabular}{cc}
\includegraphics[trim=1.5cm 10.0cm 0cm 2.0cm ,clip, scale=0.48] {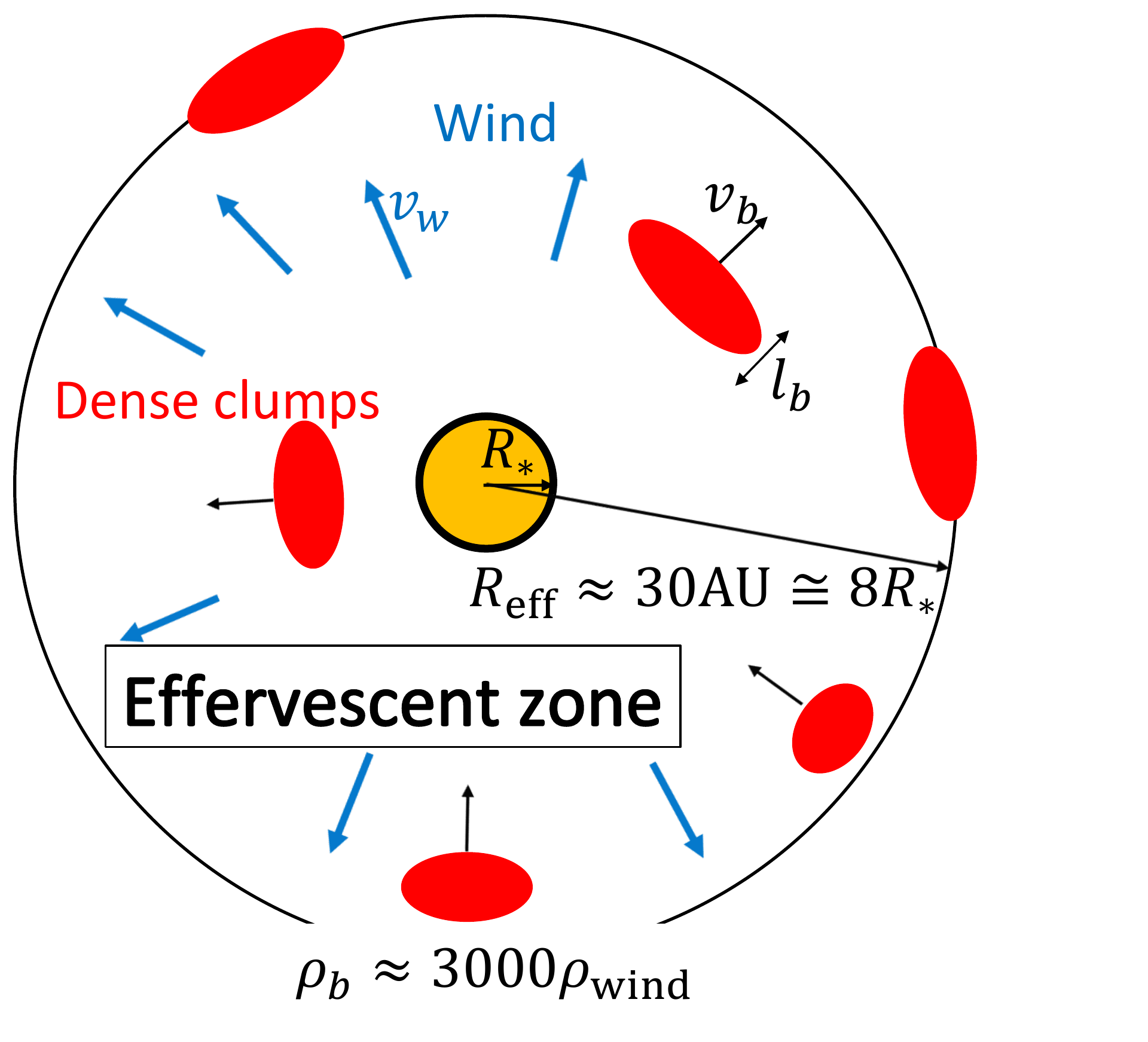} \\
\end{tabular}
\caption{A schematic drawing of the effervescent zone crudely scaled to a possible effervescent zone around the progenitor of SN~2023ixf at explosion. 
The thick-blue arrows depict the regular (escaping) wind outflowing with super-escape velocity $v_w$. The red-oval shapes represent the bound dense clumps that rise and fall within the effervescent zone. The orange sphere at the center is the RSG progenitor of SN~2023ixf of radius $R_\ast \simeq 800 R_\odot$. The outer edge of the effervescent zone is at $R_{\rm eff} \approx 30 \AU$. The typical ratio of the density of a clump to that of the regular wind at a given radius is $k_b \equiv \rho_c/\rho_{\rm wind} \approx 3000$, according to a simple model where most clumps are in a shell near $R_{\rm eff}$ and their solid angle covering fraction is $f_S \simeq 0.5$. 
}
  \label{fig:Effervescent}
    \end{figure}

I do not repeat here the derivations from \cite{Soker2021effer}, but rather only present the basic ingredients of the effervescent zone model that I present schematically in Fig. \ref{fig:Effervescent} with parameters for the progenitor of SN~2023ixf. 

The clumps in the effervescent zone are uplifted by stellar pulsations with additional uplifted forces by strong convection and possibly magnetic activity and/or rotation. The progenitor of SN~2023ixf had large-amplitude pulsations before explosion (e.g., \citealt{Kilpatricketal2023, Soraisametal2023}). The period is about $2.8 \yr$, which \cite{Kilpatricketal2023} note to be consistent with kappa-mechanism pulsations of RSG stars but with a much larger amplitude. These are the type of pulsations that can support an effervescent zone. The stellar radiation cannot accelerate the clumps to escape velocities. 
Actually, the stellar radiation is already accelerating the regular (escaping wind) to about the maximum possible mass loss rate from momentum balance 
$\dot M_{\rm wc} v_{\rm w} \simeq \eta_{\rm w} L/c$, where $v_{\rm w}$ is the terminal wind speed, $L$ the stellar luminosity, and $\eta_{\rm w}$ is the average number of times that a photon transfers its momentum to the wind in the outward radial direction. Generally $\eta_{\rm w} < 1$, but in dense and opaque winds this factor can be $\eta_{\rm w} \ga 1$.  Substituting typical values gives
\begin{eqnarray}
\begin{aligned} 
\dot M_{\rm wc} & \simeq  4 \times 10^{-5} \eta_{\rm w}
\left( \frac{L}{2 \times 10^5 L_\odot} \right)
\\ & \times 
\left( \frac{v_{\rm w}}{100 \km \s^{-1}} \right)^{-1}
M_\odot \yr^{-1} .
\label{eq:mwc1}
\end{aligned}
\end{eqnarray}
I scale the luminosity according to the inferred values of 
$L\simeq 1.3 \times 10^5 L_\odot$ \citep{Jencsonetal2023}, 
$L \simeq 1.6 \times 10^5 - 3 \times 10^5 L_\odot$ \cite{Soraisametal2023}  and $L\simeq 5.5 \times 10^4 L_\odot$ \citep{Kilpatricketal2023}.  
I scale the wind velocity for SN~2023ixf according to \cite{SmithNetal2023} who infer it to be $v_{\rm w}=115 \km \s^{-1}$.

\subsection{The density of the clumps}
\label{subsec:ClumpDensity}
Consider a clump that is ejected from the star and its density is $k_b$ times the density of the regular wind at the same radius. The assumptions \citep{Soker2021effer} are that it expands radially, i.e., its cross section facing the star varies as $A_b(r) \propto r^{2}$, and that its width (along the radial direction) $l_b$ stays constant. The forces that act on a clump are the stellar gravity (inwards; the negative radial direction), and two outwards forces, the drag by the regular wind and the radiation pressure. When the wind mass loss rate is as given by equation (\ref{eq:mwc1}) for $\eta_{\rm w}\simeq 1$ it turns out that the expression for the maximum radius inside which the acceleration is negative (otherwise the clump escapes with the wind) does not depend on whether it is optically thin or thick and it reads  \citep{Soker2021effer}. 
\begin{equation}
r_{\rm max} \simeq \frac{v_{\rm Kep}}{v_w} 
\left( \frac{l_b}{R_\ast} \right)^{1/2}    k^{1/2}_b R_\ast, 
\label{eq:rmax}
\end{equation}
where $v_{\rm Kep} = v_{\rm esc} /\sqrt{2}$ is the Keplerian velocity on the stellar surface of the RSG star and $v_{\rm esc}$ is the escape velocity. 

The escape velocity of the progenitor of SN~2023ixf is uncertain. 
\cite{Hosseinzadehetal2023} gives progenitor radius of $R_\ast \simeq 410 R_\odot$. From the luminosity and effective temperatures that studies infer the progenitor stellar radius is $R_\ast \simeq 980 R_\odot$ \citep{Jencsonetal2023} and $R_\ast \simeq 500 R_\odot$ \citep{Kilpatricketal2023}. The Mass of the progenitor is also uncertain. \cite{Jencsonetal2023} estimate the initial mass as $M_{\rm init} = 17 \pm 4 M_\odot$. On the other hand \cite{Kilpatricketal2023} argues that the luminosity they infer is consistent with $M_{\rm init} = 11 M_\odot$ and \cite{PledgerShara2023} estimate that $M_{\rm init} = 8-10 M_\odot$. \cite{Neustadtetal2023} give the range of $M_{\rm init} = 9-14 M_\odot$. \cite{Soraisametal2023} infer a value of $M_{\rm init} = 20 \pm 4 M_\odot$. I scale here with a mass at explosion of $M_\ast = 10 M_\odot$. 
Considering the long-period pulsation of $1091 \pm 71$~d \citep{Soraisametal2023}, as \cite{Kilpatricketal2023} also report, I scale the radius at explosion with 
$R_\ast \simeq 800 R_\odot$. This gives for the escape velocity and for the Keplerian velocity $v_{\rm esc}= 69 \km \s^{-1}$ and $v_{\rm Kep}= 49 \km \s^{-1}$ , respectively. 

I scale the outer radius of the compact CSM, which here is also the outer radius of the effervescent zone, with $R_{\rm eff} = R_{\rm CSM} = 30 \AU = 6445 R_\odot$ (section \ref{sec:intro}). From equation (\ref{eq:rmax}) with $r_{\rm max} = R_{\rm eff}$, I find the condition on the density factor of the clump to stay bound at the radius of the effervescent zone $R_{\rm eff}$ to be 
\begin{equation}
k_b \ga 250
\left( \frac{R_{\rm eff}}{8 R_\ast} \right)^2 
\left( \frac{v_w}{2 v_{\rm Kep}} \right)^2
\left( \frac{l_b}{R_\ast} \right)^{-1}  .
\label{eq:kbREff}
\end{equation}
Equation (\ref{eq:kbREff}) gives the lower limit on the density ratio of clumps that reach the radius of the effervescent zone and then fall back. Clumps with lower density will be dragged out by the wind at that radius. For the wind parameters used here the density of the clumps at $R_{\rm eff}$ is 
\begin{eqnarray}
\small
\begin{aligned} 
\rho_b(R_{\rm eff})  \ga & 2.5 \times 10^{-14} 
\left( \frac{\dot M_{\rm w}}{4 \times 10^{-5} M_\odot \yr^{-1}}  \right) 
\\ & \times
\left( \frac{R_\ast}{800 R_\odot} \right)^{-2} 
\left( \frac{v_{\rm w}}{100 \km \s^{-1}} \right)^{-1}
\\ & \times
\left( \frac{v_w}{2 v_{\rm Kep}} \right)^2
\left( \frac{l_b}{R_\ast} \right)^{-1}  \g \cm^{-3}.
\label{eq:RhoREff}
\end{aligned} 
\end{eqnarray}
\normalsize

The value of $l_b$ is not determined. The model assumes that very large convective cells that move outward, acting together with the pulsation modes, e.g., \cite{Freytagetal2017} and \cite{Nhungetal2023} for AGB stars, eject clumps (blobs) with sizes of $\simeq 0.2-0.5 R_\ast$. Shortly after it is formed and before it cools a clump expands in all directions due to its high pressure. For that it is scaled with $l_b=R\ast$ \citep{Soker2021effer}. 
Near the surface of the star the density of the clump is  
$\simeq (R_{\rm eff}/R_\ast)^2 \rho_b(R_{\rm eff})$ which for the scaling used here that is appropriate for the progenitor of SN~2023ixf it is $\ga 2 \times 10^{-12} \g \cm^{-3}$. This is about 2-3 orders of magnitude smaller than the density of the photosphere of RSG stars.  

\subsection{Global Properties}
\label{subsec:GlobalProperties}

The density profile of the clumps in the effervescent zone depends on the velocity distribution of the clumps when they are ejected from the star and on the initial density contrast $k_b$  
\citep{Soker2021effer}. Because the clumps slow down as they rise and accelerated as they fall back to the star, the average density is not $r^{-2}$. It can be shallower than this or steeper, depending on the initial velocity distribution. These properties span an undetermined parameter space of the effervescent zone model. 
I consider therefore global and average properties. 

Observations infer different CSM properties, depending on the type of observation and analysis.  \cite{Bostroemetal2023} infer a mean density of $5.6 \times 10^{-14} \g \cm^{-2}$. \cite{Grefenstetteetal2023} estimate a mass loss rate of $\dot M_{\rm w} \approx 3 \times 10^{-4} M_\odot \yr^{-1}$ for a wind velocity of $v_{\rm w} = 50 \km \s^{-1}$, while \cite{JacobsonGalanetal2023} estimate a value of $\dot M_{\rm w} \approx 10^{-2} M_\odot \yr^{-1}$ for $v_{\rm w} = 50 \km \s^{-1}$. The mass of the compact CSM of SN~2023ixf is in the range of $0.001-0.03 M_\odot$. I scale with $M_{\rm CSM} = 0.01 M_\odot$.

According to the effervescent zone model there is a regular wind in addition to the dense clumps. For that, the covering fraction of the clumps, which is the total solid angle the clumps cover divided by $4 \pi$, cannot be more than about half. Considering that the clumps spend a large fraction of the time near their turning point that is close to $R_{\rm eff}$ I assume a simple model where all clumps are in a shell of width $l_b$ near $R_{\rm eff}$. The demand on the covering fraction to be $f_S<0.5$ yields a constraint on the density of the clumps there to be   
\begin{eqnarray}  \small
\begin{aligned} 
\rho (R_{\rm eff}) & \ga   2.8 \times 10^{-13}
\left( \frac{M_{\rm CSM}}{0.01 M_\odot} \right)
\left( \frac{f_S}{0.5} \right)^{-1}
\\ & \times
\left( \frac{R_{\rm eff}}{30 \AU} \right)^{-2}
\left( \frac{l_b}{800R_\odot} \right)^{-1}  \g \cm^{-3} .
\label{eq:fb}
\end{aligned} 
\end{eqnarray}
\normalsize
By equations (\ref{eq:RhoREff}) and (\ref{eq:kbREff}) this implies that the density factor of the clumps is $k_b \ga 3000$. Since some clumps overlap in the solid angle they cover, the covering fraction for these parameters is actually $f_S < 0.5$. 

Consider the total volume that the clumps occupy inside the effervescent zone $V_b$. For a wind mass loss rate and velocity as the scaling of equation (\ref{eq:mwc1}) the total wind mass inside $R_{\rm eff}=30 \AU$ is $M_{\rm wind,eff} = 5.7 \times 10^{-5} M_\odot$. The relative volume the clumps occupies inside this radius is therefore
\begin{eqnarray}  \small
\begin{aligned} 
& f_V \equiv \frac{V_b}{V(R_{\rm eff})} 
\simeq 0.06 
\left( \frac{k_b}{3000} \right)^{-1}
\left( \frac{R_{\rm eff}}{30 \AU} \right)^{-1}
\\ & \times
\left( \frac{M_{\rm CSM}}{0.01 M_\odot} \right)
\left( \frac{v_{\rm w}}{100 \km \s^{-1}} \right)
\left( \frac{\dot M_{\rm w}}{4 \times 10^{-5} M_\odot \yr^{-1}} \right)^{-1} .
\label{eq:Volume}
\end{aligned} 
\end{eqnarray}
\normalsize


\section{Discussion and Summary}
\label{sec:Summary}

This study presents the effervescent zone model as an alternative scenario to the enhanced mass-loss rate phase few years before explosion as an explanation for the compact CSM around the progenitor of SN~2023ixf. A schematic drawing of the long-lived effervescent zone is in Fig. \ref{fig:Effervescent}. The motivation to propose the effervescent zone model comes from that there are no indications for outbursts in the time period of years before SN~2023ixf explosion and the large-amplitude pulsations of its progenitor. Such pulsations are expected to facilitated the formation of an effervescent zone. Previous studies of SN~2023ixf ignored the possibility of an effervescent zone. 

There are some undetermined properties of the effervescent zone model, including the density and velocity distributions of the clumps that the RSG star ejects as a result of its pulsations and envelope convection. 
For that, at this stage I limit the study to present the possible properties of individual clumps (section \ref{subsec:ClumpDensity}) and the global properties in the effervescent zone (section \ref{subsec:GlobalProperties}). Because the CSM properties of SN~2023ixf are not well determined, the quantities I derive are all scaled. These include the lower limit on the ratio of the clump density to the regular wind density (equation \ref{eq:kbREff}) and its density at the effervescent radius (equation \ref{eq:RhoREff}) in order for the clump to fall back. A stronger lower bound on the clumps' density comes from the global demand that the clumps supply the CSM mass, but that they do not cover more than about half of the solid angle in order to allow the existence of a regular (escaping) wind (equation \ref{eq:fb}). The total volume the clumps occupy within the effervescent zone for the parameters I use here is $f_V \simeq 0.06$ (equation \ref{eq:Volume}). 

The effervescent zone model has the following implications. First, as said, there is no need for a pre-explosion outburst during the several years before explosion. Secondly, the dense gas might show red-shifted and blue-shifted emission of up to several tens of $\km \s^{-1}$, and any value in between the red and blue-shifted emissions. There are a total of $N_c \approx 50$ clumps in the effervescent zone. About half dominate the  emission towards the observer. Therefore, the spectrum is a combination of these clumps, which will have one peak at the stellar velocity relative to the observer, with a spread due to Doppler shifts.   
The regular wind, however, will have stronger blue shifted emission. This might lead to a complicated spectrum. 
In particular, if the dense clumps have a small covering fraction along the line of sight to the photosphere of the ejecta then their influence on the absorption lines is small. Our next step will be to calculate the absorption and emission properties of the effervescent zone. 

For the lack of narrow blueshifted absorption lines \cite{SmithNetal2023} suggest that most of the dense CSM is not along our line of sight. They further propose that a binary interaction ejected the dense CSM into a disk or a torus in the equatorial plane, and that this disk/torus does not cross our line of sight to the supernova photosphere. They attribute the disappearance of CSM lines to a flow structure where the supernova ejecta engulf the highly non-spherical CSM. In the effervescent zone model that I suggest here for SN~2023ixf the dense clumps replace the equatorial mass ejection. The small solid angle coverage by the dense clumps, $f_S<0.5$, and the small filling fraction $f_V$ (equation \ref{eq:Volume}), suggest that the clumps might play the same role as dense equatorial disk or torus.  
 
I encourage future studies of the ejecta-CSM interaction of SN~2023ixf to consider the effervescent zone model. 

\section*{Acknowledgments}
 
 I thank an anonymous referee for useful comments.  This research was supported by a grant from the Israel Science Foundation (769/20).



\label{lastpage}

\end{document}